\documentclass[amsmath,amssymb,aps,letterpaper,prl,twocolumn]{revtex4-1}
\usepackage{graphicx}
\usepackage{color}
\usepackage{enumitem}
\usepackage{appendix}

\usepackage{picture}
\usepackage{calc}

\usepackage{subfigure}
\usepackage{dsfont}
\usepackage{color}

\usepackage[utf8x]{inputenc}

\newcommand{\ket}[1]{\mathopen{|}#1\mathclose{\rangle}}

\begin{document}

\title{Distance to integrability and complexity}

\author{Sean Casey$^{a,b}$}%
 \email{sean.casey@ibm.com}
\author{Marius de Leeuw$^b$}%
 \email{mdeleeuw@maths.tcd.ie}
\author{Tristan McLoughlin$^b$}%
 \email{tristan@maths.tcd.ie }
\author{Sergiy Zhuk $^a$}%
 \email{sergiy.zhuk@ie.ibm.com }

\affiliation{\\
\smallskip
$^a$IBM Quantum, IBM Research Europe, \\
Trinity Business School, Trinity College Dublin\\
Dublin, Ireland
\smallskip\\
$^b$School of Mathematics
\& Hamilton Mathematics Institute\\
Trinity College Dublin\\
Dublin, Ireland \\
}%

\begin{abstract}
We consider nearest-neighbour interactions on spin-1/2 chains. We find the full set of integrable models of this form 
and apply a numerical algorithm to determine the ``closest'' integrable model for a given Hamiltonian. This allows us to 
define a model-independent distance to integrability. We show that standard measures of chaos interpolate between integrable 
and chaotic values with increasing values of this distance. However, there is a large variation across models and the distance 
to integrability provides additional information not directly provided by these diagnostics.
\end{abstract}

\maketitle


\textit{Introduction} --- Perturbations of quantum integrability have been studied in great detail, but typically by deforming a given integrable model. This results in model-dependent conclusions on how the perturbation strength affects the physical behaviour. In this work, we change the direction of thinking and ask: Given a translationally invariant nearest-neighbour spin chain, how close is it to being integrable? How
does this distance relate to standard measures of integrability breaking and quantum chaos?

The key ingredient that enables us to address these questions is the classification of all integrable nearest-neighbour Hamiltonians
for a translationally invariant spin-$1/2$ system. We find 12 independent integrable models and for a given random model $M$, we can then
consider the Frobenius norm and compute the distance between $M$ and the integrable models $H$. This distance is minimised 
numerically, and gives the ``closest'' integrable model to $M$. In this way, we obtain a well-defined and scalable notion of \textit{distance to integrability}. 
 In order to deal with physically relevant models and for numerical efficiency we work with real symmetric matrices.

There are several standard diagnostics of quantum chaos which in essence measure how closely a system is described by random matrix theory (RMT) \cite{Guhr:1997ve, Haake:2010fgh, DAlessio:2015qtq}. Integrable models instead typically have Poissonian spectral statistics \cite{Montambaux:1993zz, Poilblanc:1993xz,hsu1993level} and in finite volume the transition from integrable to chaotic behaviour is generally a smooth function of the deformation strength. Studies of this crossover \cite{rabson2004crossover, santos2004integrability} typically compute such diagnostics along deformations of specific integrable models and so their conclusions inherit a dependence on the chosen starting point. We instead sample the full space of nearest-neighbour Hamiltonians by generating a
very large number of Hamiltonian densities with random real entries. For each model we numerically compute standard indicators of chaos, namely ratios of level spacings and eigenstate Shannon entropy, and find that they are non-trivially correlated with the distance to integrability. In particular, we find that the distance to integrability typically restricts the values that these diagnostics take and their rolling averages over many models interpolate between integrable and chaotic values.
 
However, the measures of chaos have a large variance and there is strong model dependence. This variance is typically largest for models that are close to integrability. This is not surprising as the transition is known to depend not only on the strength of the perturbation but also on the system size \cite{modak2014universal, modak2014finite}, the Fock-space structure of the deformation \cite{altshuler1997quasiparticle, bulchandani2021onset}, and
the existence of quasi-conserved charges \cite{Szasz-Schagrin:2021pqg, McLoughlin:2022jyt, PhysRevB.105.104302, PhysRevResearch.5.043019, vanovac2024finite, adans2026deforming}, among other properties. Chaos diagnostics are therefore not reliable indicators of how close a model is to integrability in an absolute sense and this distance to integrability provides a more robust measure of the size of the perturbation from the nearest integrable Hamiltonian. 

A particularly interesting measure is Krylov complexity, which quantifies the spreading of an initially local operator through operator space \cite{Nandy2025QuantumDynamics, Rabinovici2025KrylovComplexity}. 
Rather than uniformly slowing or accelerating this spreading, proximity to integrability broadens the range of possible values of the first 
few Lanczos coefficients, allowing both anomalously slow and anomalously fast early-time dynamics. In comparison, far from integrability, the dynamical behaviour is more 
constrained.


We attach our complete dataset to the paper so that one can find models that have specific properties (such as, distance to integrability, spectral statistics etc.) that 
are of interest and study their properties. 

\begin{figure}
    \centering
    \includegraphics[width=1.0\linewidth]{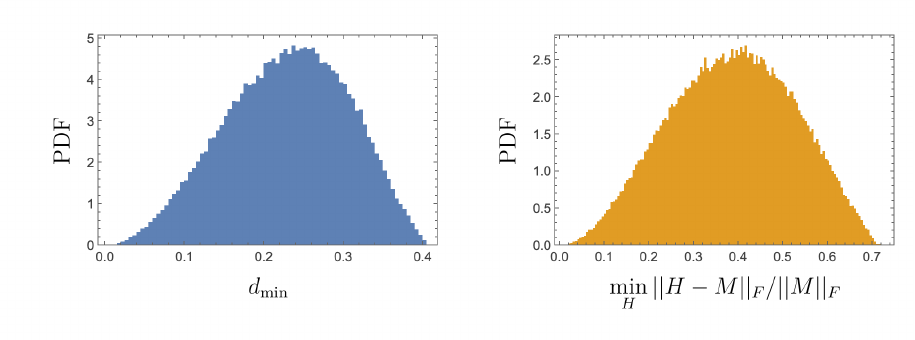}
    \caption{\textbf{Left: }Distribution of chain-length-independent distance measure for 200,000 randomly generated real symmetric Hamiltonian densities. All 
    generated densities are normalised after identity and telescoping terms are removed. \textbf{Right: }Distribution of distance to nearest integrable model
    relative to Frobenius norm of the random model on the full chain.}
    \label{fig:distance-distributions}
\end{figure}

\textit{Integrable nearest-neighbour models} --- Regular integrable models possess a tower of conserved charges, $\mathbb{Q}_n$, where
$\mathbb{Q}_n$ is an $n$-body charge. The Hamiltonian is typically taken to be $\mathbb{Q}_2= \sum_{j = 1}^L h_{j, j+1}(u)$, 
which is an operator that acts on neighbouring sites and $u$ is the so-called spectral parameter. 
Using the boost operator formalism, the rest of the charges can be recursively calculated by commuting $\mathbb{Q}_2$ with the 
so-called boost operator $\mathbb{B} $ \cite{Tetelman,HbbBoost,Loebbert:2016cdm} such that $\mathbb{Q}_{ r+1}(u) \sim [\mathbb{B}, \mathbb{Q}_r(u)]$.
This defines a tower of conserved charges for any value of the spectral parameter $u$. We can set $u=0$ and denote $\mathbb{Q}_2 := \mathbb{Q}_2(0)=\sum_j h_{j,j+1}$ and $\mathbb{G}_2 :=  \mathbb{Q}'_2(0) = \sum_j g_{j,j+1}$. From the boost operator it follows that
\begin{align}
 \mathbb{Q}_3  = -\sum_{j =1}^L& [h_{j, j+1}, h_{j+1, j+2}] +\mathbb{G}_2. \label{eq: Q3 definition}
\end{align}
Commutation of $\mathbb{Q}_2$ and  $\mathbb{Q}_3$ is a necessary and sufficient condition for integrability of $\mathbb{Q}_2$ \cite{Hokkyo:2025tgq, Sanatani:2026avv}, so 
the condition of integrability reduces to
\begin{equation}
    [\mathbb{Q}_2, \mathbb{Q}_3] = 0. \label{eq: Integrability condition}
\end{equation}
This yields a set of cubic polynomial equations in $h,g$. Following \cite{Corcoran:2023zax} we can solve these conditions
when our local Hilbert space is two-dimensional. We have carried out this classification and find that there are 12 independent families that
we list in the end matter.


\textit{Distance to integrability} --- From the full classification of integrable models, we address the problem of how to find the closest integrable model $H_M$
for any translationally invariant Hamiltonian $M = \sum_{j = 1}^L m_{j,j+1}$. In order to calculate $H_M$ we need to minimise
\begin{align}
    D^2(m) &= \min_{H}\left|\left|M - H\right|\right|_F^2,
&& ||A||_F = \sqrt{\text{Tr}(A^\dagger A)} \notag
\end{align}
over the $12$ integrable Hamiltonian densities that cover all integrable solutions.

Define $\mathcal{I}$ as the set of all integrable parameterisations shown in the end matter. Integrability is preserved by local basis transformations
and this property was exploited while solving the integrability constraint. Therefore to cover 
the most general set of integrable models we must allow for optimisation over local basis transformations. 
\begin{equation}
    D^2(m) = \min_{h \in \mathcal{I}, V\in \text{SL}(2,\mathbb{R})}\Big|\Big|\sum_{j = 1}^L (m_{j,j+1} - h^V_{j,j+1})\Big|\Big|_F^2, \label{eq: Optimisation problem}
\end{equation}
where $h^{V}_{j,j+1}:= (V\otimes V) h_{j,j+1} (V\otimes V)^{-1}$. Doing this we get an optimisation problem for the coefficients of the Hamiltonian
and the coefficients of the local basis transformation $V$. To remove the variable of the chain length $L$ we define
\begin{equation}
    d_{\min}^2(m) = \frac{1}{L 2^L} D^2 (m)
\end{equation}
which for traceless $m$ is independent of the chain length. All of the 12 integrable Hamiltonian densities in the
end matter are parameterised linearly. Hence for fixed $V$, the optimal linear parameter choices have a well-defined solution by least squares. Therefore 
the non-convex part of~\eqref{eq: Optimisation problem} is the three-parameter minimisation over the local basis transformations $V$.
Note that the closest integrable model may not necessarily be exactly symmetric.


\textit{Generating the dataset} --- In order to study how the integrable models are embedded in the space of nearest-neighbour Hamiltonians and their relation to measures of chaos, we generate
a large set of random nearest-neighbour Hamiltonian densities $m$ with entries drawn uniformly from $[-1,1]$, and then apply the above minimisation procedure to find the closest integrable model. We remove the components proportional to $\mathbb{I}$ 
and the three telescoping terms of the form $A\otimes \mathbb{I}_2 - \mathbb{I}_2 \otimes A$ \footnote{The telescoping terms vanish for translationally invariant systems with periodic boundary conditions, which is the case we consider.},
then normalise the density $m$ to unit Frobenius norm. Since we are interested in Hamiltonians with physical applications and that are diagonalisable we restrict
to symmetric matrices, and further restrict to real entries for stability and simplicity in the optimisation.

The full dataset is available at \cite{casey2026dataset}. It contains all the randomly generated Hamiltonians, the distance to the various integrable models as well as the 
values of the diagnostics that we have computed.

The first thing we can learn from our dataset is the distribution of distances of random nearest-neighbour models to their nearest integrable model, see  Fig.~\ref{fig:distance-distributions}.


\textit{Measures of chaos} --- We now look at several measures that distinguish between chaotic and integrable behaviour, and at  how they relate to a model's distance from the nearest integrable model.
\begin{figure}
	\centering
	\raisebox{-0.5\height}{\includegraphics[scale=0.23]{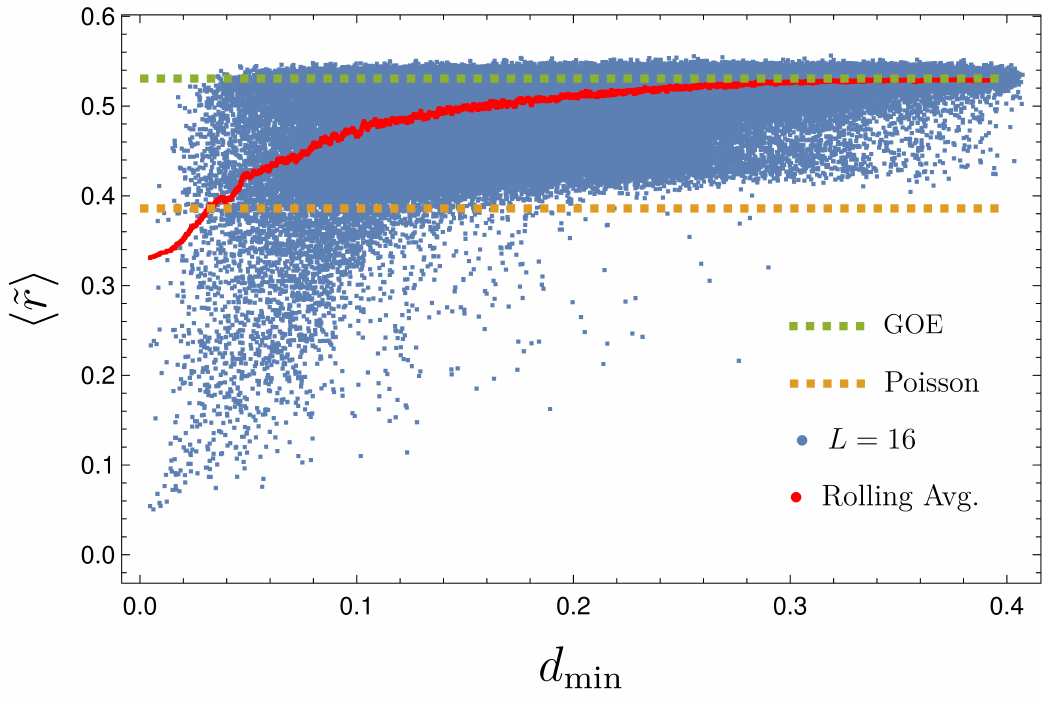}}\hspace{0.3cm}	\raisebox{-0.5\height}{\includegraphics[scale=0.23]{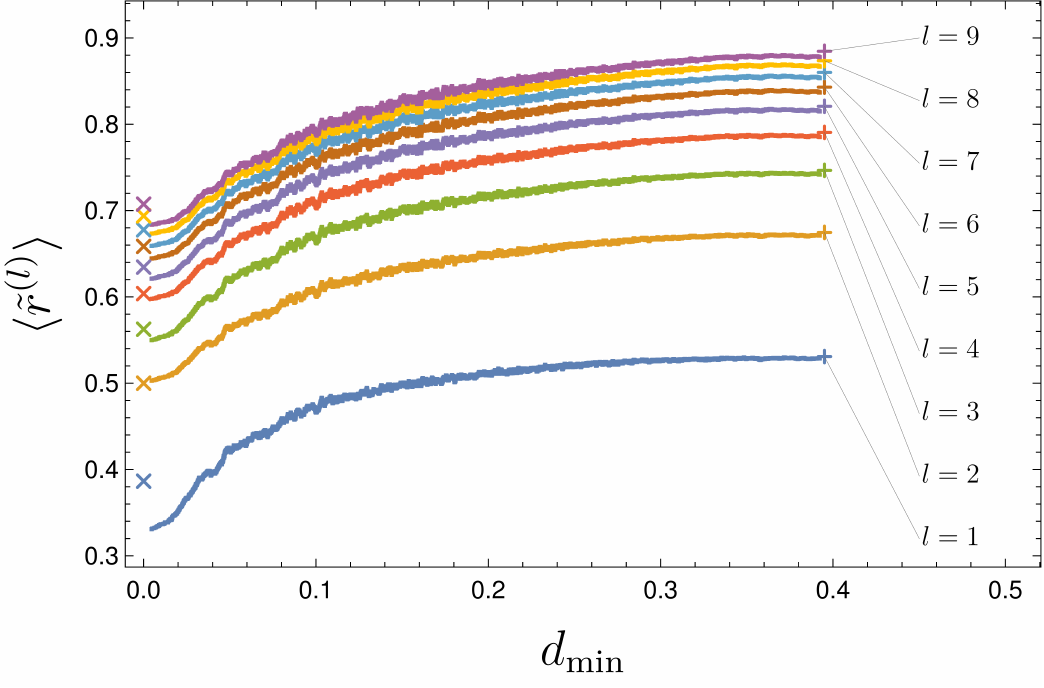}}\\
	\raisebox{-0.5\height}{\includegraphics[scale=0.23]{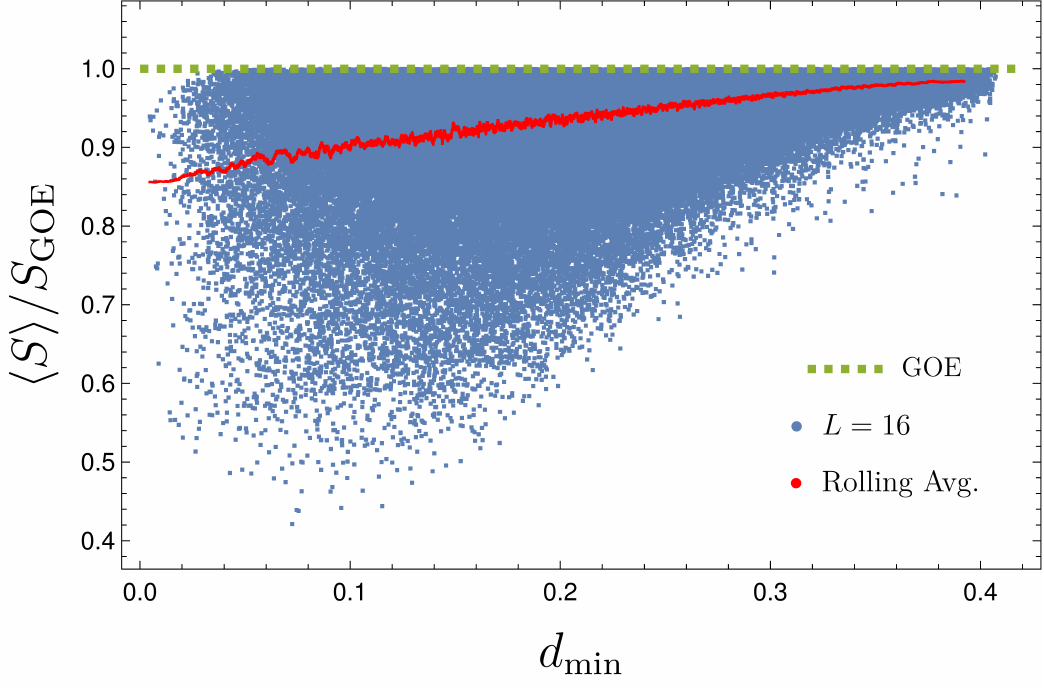}}\hspace{0.3cm}	\raisebox{-0.5\height}{\includegraphics[scale=0.23]{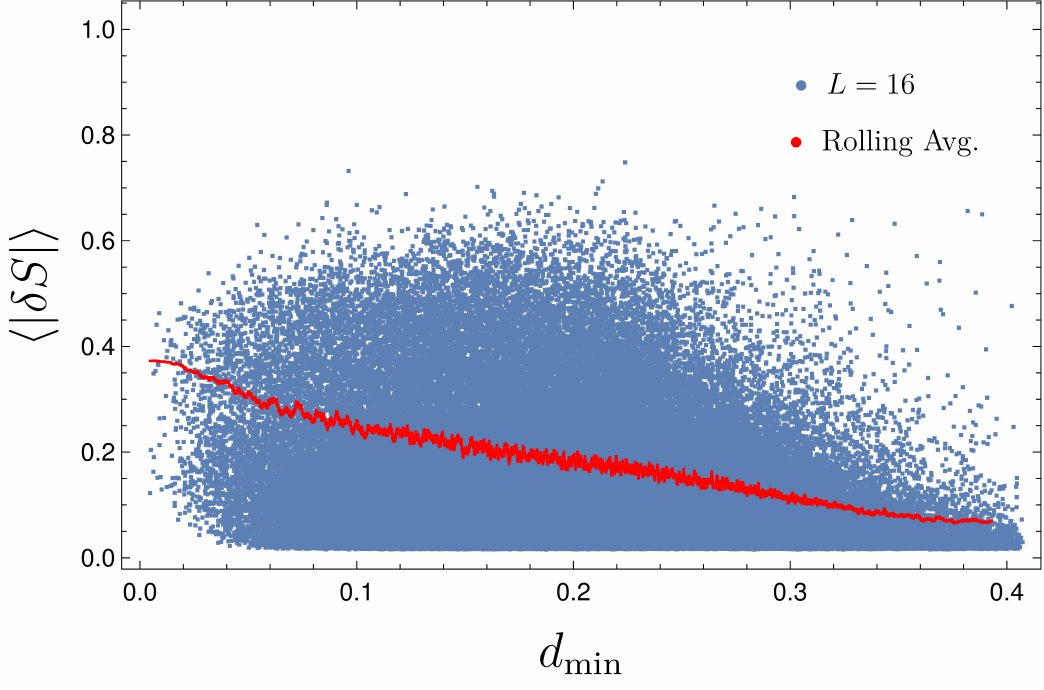}}
	\caption{\textbf{Top Left:} $\langle \tilde{r}\rangle$ for randomly generated $L=16$ spin-chain Hamiltonians as a function of $d_{\text{min}}$. 
		 \textbf{Top Right:} Rolling average of generalised ratio statistics $\langle \tilde{r}^{(l)}\rangle $ with $l=1, \dots, 9$. X's at the left and crosses at the right ends of each curve mark the Poisson and GOE values, respectively.
		  \textbf{Bottom Left:} Shannon entropy,  in the spin basis, averaged over eigenvectors with energies in the central 70\% of the spectrum. \textbf{Bottom Right:} Mean absolute difference, $\langle|\delta S|\rangle$, between entropies of eigenstates with neighbouring energies. }
	\label{fig:mchaos}
\end{figure}

\textit{Spectral statistics} --- The presence or absence of quantum chaos in a particular system is most commonly diagnosed through the use of measures based on the spectrum of energy eigenvalues \cite{Bohigas:1983er}. The distribution of spacings $s_i=E_{i+1}-E_i$, where $E_i$, $i=1, 2, \dots , D$, are the energy eigenvalues in ascending order, is predicted by random matrix theory (RMT) to be approximately Wigner--Dyson for chaotic models, reflecting the underlying property of level repulsion. On the other hand, integrable models typically have a Poisson distribution \cite{berry1977level}.
A convenient quantity that measures how closely a given spectrum approaches that of a typical chaotic system is the $r$-statistic, which is defined in terms of ratios of consecutive level spacings $r_i=s_i/s_{i-1}$ as \cite{Oganesyan:2007wpd}, $
	\tilde{r}_i=\min \left(r_i, \frac{1}{r_i}\right)$.
For real, symmetric Hamiltonians -- corresponding to the RMT Gaussian Orthogonal Ensemble (GOE) -- a Wigner-like surmise for the distribution \cite{Atas:2013gvn} predicts a mean value $\langle \tilde{r}\rangle_{\text{GOE}}=0.5359$, which is close to the numerically determined value of $\simeq 0.5307$, while the Poisson value is $\langle \tilde{r}\rangle_{\text{P}}=0.3863$.  We numerically compute the $r$-statistic for each Hamiltonian in our dataset by fixing the spin-chain length $L$ and directly diagonalising \footnote{As our models are translationally invariant, lattice momentum is a conserved quantum number such that the Hamiltonian comprises independent blocks from which we choose the zero-momentum sector since it is generally the largest. As states at the edges of the spectrum typically do not become ergodic, we clip each spectrum by removing 15\% of the levels from each edge. }.
In Fig.~\ref{fig:mchaos} (Top Left) there is a clear trend: models with small $d_{\text{min}}$ have $\langle{\tilde r}\rangle$  closer to the Poisson value, while models with larger $d_{\text{min}}$ generally approach the GOE chaotic value. This relation is made particularly clear by smoothing the data \footnote{This smoothing is done by sorting the points by $d_{\text{min}}$ and taking a rolling average over 500 data points.}, which produces an almost monotonic curve starting close to the Poisson value and rising to the RMT prediction \footnote{One additional feature is a number of data points significantly below the integrable Poisson value. Examining a sample of these points, we find that they are numerically close to non-interacting models with highly clustered spectra.}. 
The transition is relatively quick, and already for $d_{\text{min}}\gtrsim 0.07$, more than half of these $L=16$ models have $\langle{\tilde r}\rangle > 0.459$, the midpoint between the Poisson and  GOE values. The transition occurs more quickly with increasing $L$ and it is generally expected that in the thermodynamic limit any finite deformation will result in a chaotic spectrum. 
 More generally, while the distance to integrability and smoothed values of $\langle{\tilde r}\rangle$ show a clear relation, there is large variance, with chaotic values of $\langle{\tilde r}\rangle$ occurring even at small $d_{\text{min}}$. This reflects the fact that different directions in deformation space, away from different integrable models, can result in quite distinct physical behaviour.

RMT makes predictions for chaotic spectra that go beyond neighbouring energy levels and extend to arbitrary correlators of eigenvalues.
Longer-range correlations can be probed without unfolding the spectrum by generalising the $r$-statistic to ratios of $l$-th neighbour spacings~\cite{Tekur:2018nme}: $\tilde{r}^{(l)}_i=\min \left(r^{(l)}_i, 1/r^{(l)}_i\right)$ with $r^{(l)}_i=(E_{2l+i}-E_{l+i})/(E_{l+i}-E_i)$ \cite{Tekur:2018nme} \footnote{This is the non-overlapping generalisation; an overlapping version was introduced earlier in~\cite{Atas:2013gvn, Atas:2013gns} and gives qualitatively similar results for the models considered here.}. In Fig.~\ref{fig:mchaos} (Top Right) we plot smoothed curves of $\langle {\tilde r}^{(l)}\rangle $ as a function of $d_{\text{min}}$ for the $L=16$ spin chain with $l=1,\dots,9$. The same general trend can be seen for all values of $l$, with the data starting close to the Poisson value and then increasing almost monotonically to the GOE result \footnote{We numerically compute the Poisson and GOE limiting values by averaging over an ensemble of 1,500 realisations of the Rosenzweig--Porter model with $D=4{,}116$ for integrable and chaotic values of the interpolating parameter.}.


\textit{Eigenstate observables} --- While spectral observables are the most common indicators, the structure of energy eigenvectors also plays an important role in understanding the onset of chaos. The statistical properties of  eigenvectors have been extensively studied, and chaotic systems are expected to have eigenstates that are maximally delocalised \cite{Berry:1977wpp, zelevinsky1996nuclear, Borgonovi:2016nrx}. A common measure of this delocalisation is the information (or Shannon) entropy of a given eigenstate with respect to a particular basis \footnote{ 
	One convenient choice of basis for the zero-momentum sector is the computational or spin basis obtained by symmetrising each spin configuration over lattice translations. }. That is, given a basis $\ket{e_\alpha}$, $\alpha=1, \dots, D$, in which each  eigenstate can be decomposed as $\ket{E_i}=\sum_{\alpha=1}^D c^\alpha_i \ket{e_\alpha}$, the Shannon entropy is $
	S_i=-\sum_{\alpha=1}^D |c_i^\alpha|^2 \ln |c_i^\alpha|^2$.
For completely localised states, $|c_{i}^\alpha|=\delta^{\alpha\alpha_0}$ for some $\alpha_0$, the entropy is zero, while for GOE matrices in a generic basis the ensemble-averaged entropy is $S_{\rm GOE}\simeq \ln( 0.482 D)$. 
Eigenstates in typical integrable spin chains can be viewed as generalised plane waves and so are also delocalised over the spin basis with a non-vanishing entropy. Nonetheless, the states are structured and the entropy is generally lower than that of chaotic models which are, at least towards the middle of the spectrum, close to the RMT result. This can be seen in Fig.~\ref{fig:mchaos} (Bottom Left), where models close to integrability generally have lower mean entropy than those with larger $d_{\text{min}}$. Perhaps even more characteristic of the integrable-to-chaotic transition is a decrease in the variation of entropy of eigenstates; with increasing chaos, the entropy more closely approximates a smooth function of the energy. This can be characterised by the absolute value of differences of entropies for states with neighbouring energies $|\delta S|=|S_{i+1}-S_{i}|$. In Fig.~\ref{fig:mchaos} (Bottom Right) it can be seen that the mean of $|\delta S|$ generally decreases with increasing $d_{\text{min}}$, consistent with the general pattern. However, as seen for the spectral statistics, there is a large variance across models. This is particularly true at small $d_{\text{min}}$, where models with large mean entropy and small $\langle |\delta S|\rangle$ are relatively common.


\textit{Reshetikhin condition} --- Failure of a Hamiltonian to commute with its respective $\mathbb{Q}_3$ implies non-integrability in the Yang-Baxter framework. 
Therefore $||[M, \mathbb{Q}_3]||_F$ could serve as a lower-cost indicator of distance to integrability \cite{Grabowski}.
Given $M$ the candidate for the next charge is of the form shown in~\eqref{eq: Q3 definition}. 
The commutation failure can then be quantified by $\min_g||[M,\mathbb{Q}_3]||_F$. For fixed $m$ this is quadratic in $g$ and solved by least squares.
\begin{figure}
     \centering
    \includegraphics[width=0.9\linewidth]{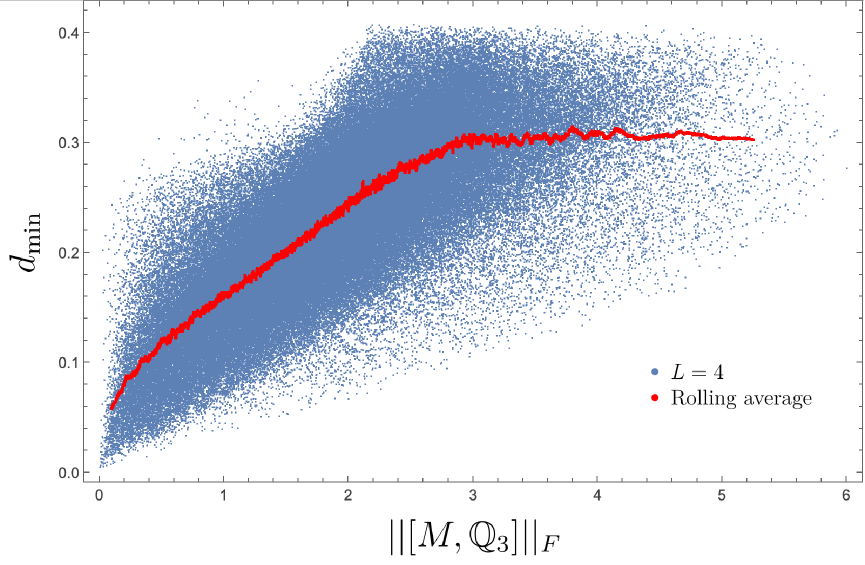}
    \caption{Plot of $d_{\min}$ versus the minimised commutator norm $\min_g||[M,\mathbb{Q}_3]||_F$ for $L=4$. The red line is a rolling average over 500 points.}
    \label{fig:Reshetikhin}
\end{figure}
From Fig.~\ref{fig:Reshetikhin}, the commutation failure is clearly positively correlated with the distance to integrability but it seems to saturate at some point.
Models with relatively small $d_{\min}$ can have comparatively large $||[M,\mathbb{Q}_3]||_F$, indicating 
that the minimum commutator norm is not a reliable proxy for deviation from integrability.


\textit{Krylov complexity} --- Krylov complexity quantifies the growth of an operator under Heisenberg evolution \cite{Nandy2025QuantumDynamics, Rabinovici2025KrylovComplexity}. Given a Hamiltonian $H$, the Liouvillian superoperator $\mathcal{L}(\cdot)=[H,\cdot]$ 
acts on operator space equipped with the Hilbert-Schmidt inner product $(A|B)=\frac{1}{2^L}\text{Tr}(A^\dagger B)$. Starting from a normalised, Hermitian,
spatially local seed operator $|O_0)$, the Lanczos algorithm \cite{Viswanath1994RecursionMethod} generates an orthonormal basis $\{|O_n)\}$ via
\begin{align}
|A_n) &= \mathcal{L}|O_{n-1}) - b_{n-1}|O_{n-2}), \nonumber\\
b_n &= \sqrt{(A_n|A_n)}, \qquad
|O_n) = \frac{|A_n)}{b_n}.
\end{align}
In this basis, $\mathcal{L}$ is represented by the tridiagonal Krylov
 Hamiltonian whose only non-zero entries are the Lanczos coefficients $b_n$.
The time-evolved operator is given by $|O(t))=e^{i\mathcal{L}t}|O_0)=\sum_n\phi_n(t)|O_n)$, and the Krylov complexity is given by quantifying the spreading 
throughout the operator basis as
\begin{equation}
C_K(t) = \sum_{n} n\,|\phi_n(t)|^2.
\end{equation}
For consistency, it is important that $O_0$ is traceless. 
A non-zero trace would introduce an identity component that neither 
evolves nor contributes to the complexity, artificially suppressing $C_K$.

Signatures distinguishing integrable and chaotic dynamics appear in the Lanczos coefficients \cite{Parker:2019ugh, PhysRevB.102.085137, PhysRevE.106.014152}. 
Previous studies have examined the late-time saturation of $C_K(t)$ \cite{Rabinovici2021Operator,Rabinovici2022Krylov,Alishahiha2025Krylov,Barbon2019OperatorComplexity} and Lanczos-sequence 
statistics \cite{Balasubramanian2025KrylovChaos,Hashimoto2023Krylov}. Here, we ask whether proximity 
to integrability directly constrains early-time growth, which is dominated by the first few Lanczos coefficients.
The first coefficient also bounds the growth at all times through
\begin{equation}
    |\partial_t C_K (t)| \le 2 b_1 \Delta \mathcal{K}, \quad \mathcal{K} = \sum_n n |O_n )(O_n|,
\end{equation}
where $\Delta \mathcal{K}^2 = \langle \mathcal{K}^2 \rangle -  \langle \mathcal{K} \rangle^2$ \cite{Hornedal2022Ultimate}. 
Since the early-time dynamics depends strongly on the seed operator \cite{hsvm-w849, PhysRevE.107.024217}, we characterise its possible range through the extremal values:
\begin{equation} 
    b_j^{\min}:=\min_{O_0}b_j(M,O_0),\qquad b_j^{\max}:=\max_{O_0}b_j(M,O_0), \label{eq: extremal lanczos}
\end{equation}
for $b_1, b_2$ and $b_3$. For $b_1$, the solution of~\eqref{eq: extremal lanczos} reduces to a small eigenvalue problem; for $b_2$ and $b_3$, however, 
it is non-convex, and heuristic global search is required.
\begin{figure}
     \centering
    \includegraphics[width=1.0\linewidth]{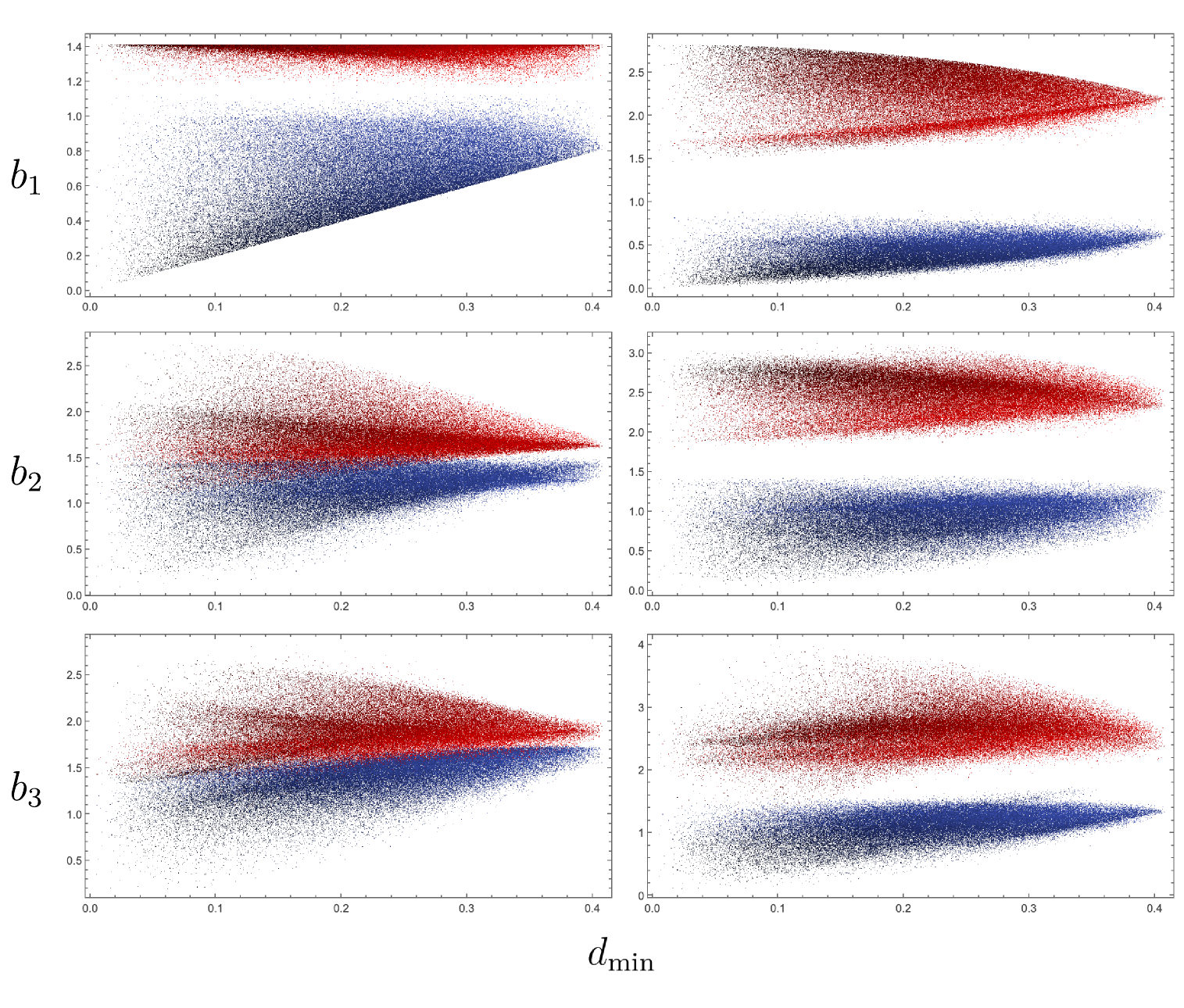}
    \caption{\textbf{Left column:} Extremal Lanczos coefficients ($b_1, b_2, b_3$) for single-site Hermitian seed operators as a 
    function of the distance to integrability. \textbf{Right column:} Corresponding extremal Lanczos coefficients for two-site 
    Hermitian seed operators. Both $b_j^{\min}$ (blue) and $b_j^{\max}$ (red) are obtained by optimising over the initial seed operator. 
    Scatter plot points are colour-coded by the local non-commutativity measure $T(m)$, where darker points are more trivial, to separate the effects of proximity 
    to integrability from simple behaviours due to local commutativity.}
    \label{fig:KrylovComplexityb123}
\end{figure}
Figure~\ref{fig:KrylovComplexityb123} shows $b_j^{\min}$ and $b_j^{\max}$ for single site initial seeds and two site initial seeds versus the distance to integrability.
Each point is colour-coded by the local non-commutativity measure $T(m)=\left\|[m_{12},m_{23}]\right\|_F$. When $T(m) = 0$, the time evolution factorises, so $T(m)$
helps to distinguish effects associated with proximity to integrability from simple behaviour caused by local commutativity.

The central result is that the distance to integrability does not uniformly slow or speed up the initial operator growth. Instead proximity permits
a broader range of both slow and fast evolutions. Interestingly, the models that are far from integrability have tighter distributions and
hence, more homogeneous early-time behaviour. Therefore, although the early-time complexity does not generally distinguish between chaotic and integrable models, the distance to 
integrability is \emph{correlated} with a constraint on the possible early-time dynamics. 
Saturation of the apparent bounds seems to be partly associated with models that are close to locally commuting. However, particularly for $b_2$ and $b_3$, the broadening persists at appreciable $T(m)$, suggesting an additional 
association with non-trivial integrability.

\textit{Conclusions} --- In this letter we found the closest integrable model for a given nearest-neighbour Hamiltonian on a spin-$1/2$ chain and we used
this to create a well-defined distance to integrability. We then studied how this correlates with standard diagnostics of chaos such
as spectral statistics and eigenstate entropy. We found that the distance to integrability constrains the values these quantities can take, and that their 
average values interpolate between the integrable and chaotic regimes. We also found constraints on the values of
the Lanczos coefficients which affect the Krylov complexity. 

There are several new directions that can be investigated. First, it would be interesting to extend our analysis to 
complex Hermitian densities. Second, the properties of the 12 different integrable models could be individually examined to uncover finer structure in our results. Third, it would also be interesting to explore whether the closest 
integrable model can be used to efficiently compute physical quantities of the original model (see \textit{e.g.} \cite{Hutsalyuk:2025zmi}). Finally, the distance depends on the choice of norm. It could be interesting to investigate norms other than the Frobenius one and see how this affects the correlations.

The dataset that accompanies this letter provides a vast set of nearest-neighbour models that can be implemented on a quantum
circuit or potentially in a condensed matter experiment. The models can be screened for different properties and a model 
of particular interest can be chosen and studied. 

We would also like to point out that the dataset can be used to devise new measures of chaos and integrability. Any quantity
that can be assigned to a Hamiltonian can easily be correlated with distance to integrability and all the other measures
that were covered in this letter. 



\begin{acknowledgments}
\paragraph{Acknowledgements.}

We would like to thank B. Pozsgay, B. Creed, P. Claeys and S. Demulder
for useful discussions and comments.
MdL is  supported by ERC-2022-CoG-FAIM 101088193. TMcL thanks the CERN Theory Department for their hospitality while this work was being completed.

\end{acknowledgments}

\bibliography{lettershortened}

\appendix
\onecolumngrid
\section*{End Matter}
\twocolumngrid

\label{Integrable Hamiltonian Densities}

\textit{Integrable Densities} --- From~\eqref{eq: Integrability condition}, the integrability condition to be solved to find all nearest-neighbour integrable models is
\begin{align}
     \Bigg[\sum_{j = 1}^4 h_{j, j+1},& -\sum_{l =1}^4 [h_{l, l+1}, h_{l+1, l+2}] + \sum_{j = 1}^4 g_{j, j+1}\Bigg] = 0, \label{eq: Integrability equation}
\end{align}
where we drop the dependence on $u$ and replace the derivative of the Hamiltonian with
 $\mathbb{G}_2 = \sum_{j = 1}^4 g_{j, j+1}$. The integrability of $H$ only
  requires the existence of some $\mathbb{G}_2$, but we do not need to solve the 
  differential equations to parameterise them explicitly since we are only interested
  in optimising over $H$. Each Hamiltonian density $h$, which is embedded in the full space via $h_{j,j+1} = \mathbb{I}\otimes ... \otimes\mathbb{I}\otimes h \otimes \mathbb{I}\otimes ... \otimes\mathbb{I}$, is a $4\times 4$ 
  matrix decomposed in the following basis:
\begin{align*}
h &= 
h_1 \mathbb{I} \otimes \mathbb{I} 
+ h_2 (\sigma_z \otimes \mathbb{I} - \mathbb{I} \otimes \sigma_z) 
+ h_3 \sigma_+ \otimes \sigma_- \\
&+ h_4 \sigma_- \otimes \sigma_+ 
+ h_5 (\mathbb{I} \otimes \sigma_z + \sigma_z \otimes \mathbb{I}) \\
&+ h_6 \sigma_z \otimes \sigma_z 
+ h_7 \sigma_+ \otimes \sigma_+ 
+ h_8 \sigma_- \otimes \sigma_- \\
&+ h_9 (\sigma_- \otimes \mathbb{I} + \mathbb{I} \otimes \sigma_-) 
+ h_{10} (\sigma_- \otimes \mathbb{I} - \mathbb{I} \otimes \sigma_-) \\
&+ h_{11} (\sigma_+ \otimes \mathbb{I} + \mathbb{I} \otimes \sigma_+) 
+ h_{12} (\sigma_+ \otimes \mathbb{I} - \mathbb{I} \otimes \sigma_+) \\
&+ h_{13} (\sigma_- \otimes \sigma_z + \sigma_z \otimes \sigma_-) 
+ h_{14} (\sigma_- \otimes \sigma_z - \sigma_z \otimes \sigma_-) \\
&+ h_{15} (\sigma_+ \otimes \sigma_z + \sigma_z \otimes \sigma_+)
+ h_{16} (\sigma_+ \otimes \sigma_z - \sigma_z \otimes \sigma_+),
\end{align*}
with $g$ decomposed in the same basis. 

Following \cite{Corcoran:2023zax}, by using the preservation of integrability under shifting by $\mathbb{I}$,
removing telescoping terms $A \otimes \mathbb{I} - \mathbb{I} \otimes A$, rescaling the Hamiltonian density by a constant and 
in particular local basis transformations of the form $h \mapsto (V\otimes V)h (V\otimes V)^{-1}$ we simplify~\eqref{eq: Integrability equation} to find the following $12$ integrable families:
\begin{widetext}
\setlength{\arraycolsep}{3pt}
\begin{gather*}
\begin{pmatrix}
h_6 & h_{11}+h_{15} & h_{11}-h_{16} & 0 \\
0 & -h_6 & 0 & h_{11}+h_{16} \\
0 & 0 & -h_6 & h_{11}-h_{15} \\
0 & 0 & 0 & h_6
\end{pmatrix}, \qquad
\begin{pmatrix}
h_5+h_6 & 0 & 0 & 0 \\
0 & -h_6 & h_3 & 0 \\
0 & h_4 & -h_6 & 0 \\
0 & 0 & 0 & -h_5+h_6
\end{pmatrix},\\[6pt]
\begin{pmatrix}
-h_0 & 0 & 0 & 0 \\
h_0+h_3 & 0 & h_3 & 0 \\
h_0+h_4 & h_4 & 0 & 0 \\
0 & h_0-h_4 & h_0-h_3 & h_0
\end{pmatrix}, \qquad
\begin{pmatrix}
h_6 & -h_0+h_9 & h_0+h_9 & 0 \\
h_0+h_9 & -h_6 & 0 & -h_0+h_9 \\
-h_0+h_9 & 0 & -h_6 & h_0+h_9 \\
0 & h_0+h_9 & -h_0+h_9 & h_6
\end{pmatrix},\\[6pt]
\begin{pmatrix}
h_6 & h_{11}+h_{13}-h_9 & -h_{11}+h_{13}-h_9 & 0 \\
h_{11}+h_{13}+h_9 & -h_6 & 0 & h_{11}-h_{13}-h_9 \\
-h_{11}+h_{13}+h_9 & 0 & -h_6 & -h_{11}-h_{13}-h_9 \\
0 & h_{11}-h_{13}+h_9 & -h_{11}-h_{13}+h_9 & h_6
\end{pmatrix}, \qquad
\begin{pmatrix}
h_6 & h_{15} & h_{15} & 0 \\
h_{13} & -h_6 & h_3 & -h_{15} \\
h_{13} & h_3 & -h_6 & -h_{15} \\
0 & -h_{13} & -h_{13} & h_6
\end{pmatrix},\\[6pt]
\begin{pmatrix}
\dfrac{h_3}{2} & h_{14} & h_9 & 0 \\
h_9 & -\dfrac{h_3}{2} & h_3 & h_{14} \\
h_{14} & h_3 & -\dfrac{h_3}{2} & h_9 \\
0 & h_9 & h_{14} & \dfrac{h_3}{2}
\end{pmatrix}, \qquad
\begin{pmatrix}
h_6 & h_{16} & -h_{16} & 0 \\
h_{14} & -h_6 & 0 & h_{16} \\
-h_{14} & 0 & -h_6 & -h_{16} \\
0 & h_{14} & -h_{14} & h_6
\end{pmatrix}, \qquad
\begin{pmatrix}
\dfrac{h_3}{2} & 0 & 0 & 0 \\
h_{14}+h_9 & -\dfrac{h_3}{2} & h_3 & 0 \\
-h_{14}+h_9 & h_3 & -\dfrac{h_3}{2} & 0 \\
0 & h_{14}+h_9 & -h_{14}+h_9 & \dfrac{h_3}{2}
\end{pmatrix},\\
\begin{pmatrix}
\dfrac{h_3}{2} & 0 & 0 & 0 \\
-h_{14}+h_9 & -\dfrac{h_3}{2} & h_3 & 0 \\
h_{14}+h_9 & h_3 & -\dfrac{h_3}{2} & 0 \\
h_8 & -h_{14}+h_9 & h_{14}+h_9 & \dfrac{h_3}{2}
\end{pmatrix}, \qquad
\begin{pmatrix}
0 & -h_{16} & h_{16} & 0 \\
-h_{14} & 0 & 0 & -h_{16} \\
h_{14} & 0 & 0 & h_{16} \\
h_8 & -h_{14} & h_{14} & 0
\end{pmatrix}, \qquad
\begin{pmatrix}
2h_5 & 0 & 0 & 0 \\
0 & 0 & h_3 & 0 \\
0& -h_3 & 0 & 0 \\
h_8 & 0& 0 & -2h_5
\end{pmatrix}.
\end{gather*}
\end{widetext}
We assume $h_8\neq 0$ for final three models.

\textit{Optimisation over integrable parameterisations} --- The above integrable solutions are in a fixed gauge achieved by tuning the basis transformation $V$ within $\text{SL}(2, \mathbb{R})$ to simplify the integrability equations.
To truly span the set of integrable models, we must allow for $\text{SL}(2, \mathbb{R})$ transformations on top of the
fixed gauge densities in the previous section. Thus, the most general integrable models are 
$h \rightarrow h^V =  (V\otimes V)(h + h_1 \mathbb{I}_4)(V\otimes V)^{-1}$. For the full length $L$ chain, we look to minimise 
\begin{align}
    D^2(m) = \min_{\substack{V\in \text{SL}(2, \mathbb{R})\\ h \in \mathcal{I}}}\Big|\Big|&\sum_{j = 1}^L \big(m_{j,j+1}- h^V_{j,j+1}\big)\Big|\Big|_F^2, \label{eq: optimisation problem full}
\end{align}
with respect to $V$ and the parameters $\{h_k\}$ of $h \in \mathcal{I}$. Decomposing the Hamiltonian density into basis matrices
\begin{align}
    h &= B_0 + \sum_{k} h_k B_k \notag\\
    \implies& (V\otimes V) h (V\otimes V)^{-1} = A_0(V)+ \sum_{k} h_k A_k(V)
\end{align}
where $B_0$ is a constant part of the base density and $A_k(V) = (V\otimes V)B_k(V\otimes V)^{-1}$. From the definition of the Frobenius
norm, $|| O||_F^2 := \text{Tr}(O^\dagger O)$, in a nearest-neighbour operator the norm simplifies as follows
\begin{align}
    &\frac{1}{2^L}\big|\big|\sum_{j = 1}^L (x_{j,j+1})\big|\big|_F^2 = \frac{L}{4}\text{Tr}(x^\dagger x) + \frac{L(L-3)}{16}|\text{Tr}(x)|^2 \notag\\
    &+ \frac{L}{8}\text{Tr}[(x^\dagger \otimes \mathbb{I}_2)( \mathbb{I}_2\otimes x)] 
    + \frac{L}{8}\text{Tr}[( \mathbb{I}_2\otimes x^\dagger)(x \otimes \mathbb{I}_2)]  
    . \label{eq: Full Space Frobenius Norm}
\end{align}
Here we see that for traceless $x$, the Frobenius norm squared scaled by $(L2^L)^{-1}$ is independent of the chain length. Now define the weighted Frobenius inner product from \eqref{eq: Full Space Frobenius Norm}
\begin{align}
    &\langle X,Y\rangle_L = \frac{L}{4}\text{Tr}(X^\dagger Y) + \frac{L}{8}\text{Tr}[(X^\dagger \otimes \mathbb{I}_2)( \mathbb{I}_2\otimes Y)]\notag \\
    &+ \frac{L}{8}\text{Tr}[( \mathbb{I}_2\otimes X^\dagger)(Y \otimes \mathbb{I}_2)] 
    + \frac{L(L-3)}{16}{\text{Tr}(X)^*\text{Tr}(Y)}.\label{eq: frobenius inner product}
\end{align}
For fixed $V$,~\eqref{eq: optimisation problem full} is proportional to
\begin{equation}
   \phi_L(\{h_k\};V) =  \langle R - \sum_k h_k A_k(V),  R - \sum_l h_l A_l(V)\rangle_L,
\end{equation}
with $R = m - A_0(V)$. In matrix form 
\begin{align}
    \phi_L(\{h_k\};V) = c(V) &- \vec{h}^\dagger g(V) - g(V)^\dagger \vec{h} \notag\\
    &+\vec{h}^\dagger G(V) \vec{h},
\end{align}
with 
\begin{align*}
    G_{kl}(V) &= \langle A_k(V), A_l(V)\rangle, \quad g_{k}(V) = \langle A_k(V), R(V)\rangle, \\
    c(V) &= \langle R(V),R(V) \rangle.
\end{align*}
The optimal coefficients satisfy the normal equations
\begin{equation}
    G(V) h_*(V) = g(V).  \label{eq: normal equations}
\end{equation}
Therefore the overall objective becomes
\begin{align}
    \min_V \Phi_L(V) = \min_V\big( c(V) &- \vec{h}_*^\dagger g(V) - g(V)^\dagger \vec{h}_* \notag\\
    &+\vec{h}_*^\dagger G(V) \vec{h}_*\big). \label{eq: projected full space objective}
\end{align}

\twocolumngrid
\end{document}